\documentstyle[preprint,revtex]{aps} 
\def\al#1{\mbox{$\alpha_ #1 ^{-1}$}}

\begin{document}

\draft

{
\begin{flushright}
LAEFF--93/014\\
October 1993
\end{flushright}
}
\begin{title}
 Influence of Light and Heavy Thresholds \\ on Susy  Unification \\
\end{title}
\author{M. Bastero--Gil\cite{aa1},
J.P\'erez--Mercader\cite{aa2}}
 \begin{instit}
Laboratorio de Astrof\'{\i}sica Espacial y F\'{\i}sica
Fundamental\\
Apartado 50727\\
28080 Madrid
\end{instit}
\begin{abstract}
 In this paper we study and compare susy unification using
two different approaches in order to take into account the effect of light
particle thresholds on the evolution of gauge couplings: the step--function
approximation, on the one hand, and a mass dependent procedure, which gives a
more accurate description of the dependence of the results on the masses,
on the other. We also include the effect of heavy thresholds, when $SU(5)$ is
chosen as the unifying group. We find that the mass--dependent procedure
excludes scenarios where all susy masses are below $1\;TeV$, and favors a value
of $\alpha_3(m_Z)$ near its upper experimental bound, contrary to the results
obtained with the step--function approximation. We underline the dependence of
the results on the procedure chosen to deal with light thresholds.
\end{abstract}
\begin{center} \bf
{\sl Submitted to Physics Letters B}
\end{center}
\narrowtext
\vspace{.2in}
\pagebreak


 Experimental data on $\sin^2\theta$ and
$\alpha_e$  measured at $m_Z$ can be used to test supersymmetric
grand unification
theories (SGUT). Use of 2--loop RGE's to go from $M_X$
(the unification scale) down to $m_Z$, gives a value for
$\sin^2\theta$ in  good agreement with experiment \cite{ellis1}.
 It  can also be
checked  that when the three gauge couplings \al{3},\al{2} and \al{1} of
$SU(3)\times SU(2) \times U(1)$ run from low to high energies, they become
equal
at a scale $M_X$  high enough to satisfy experimental lower limits on proton
decay \cite{amaldi,tsusy}. Further work has been done on this problem,
making use not only of limits on
proton decay \cite{gutpd,hisano}, but of cosmological arguments on the relic
abundance of the light susy particle (LSP) \cite{reden} to constrain
the susy spectrum.  In general, the
susy spectrum obtained from this scenario is  below 1 TeV, and
available in the next generation of particle accelerators. Their detection
would be the best test of susy theories.

In studying SGUT's one has to deal with both  light thresholds (associated
with the susy masses), and  heavy thresholds (associated with the heavy masses)
arising from the specific  unification group $G$. When supersymmetry is
broken at the Planck scale by a ``hidden" sector, the large number of susy
masses can be determined at the weak scale in terms of a small number of
parameters at the unification scale: $m_{1/2}$ (universal gaugino
masses), $m_0$ (universal scalar masses), $A$ (cubic scalar couplings)
and $m_4$ (bilinear scalar coefficient) \cite{susy}. On the other hand,
for the minimal
choice $G=SU(5)$, there are three basic heavy mass parameters: $M_V$
(heavy gauge bosons), $M_{\Phi}$ (heavy colored scalars), and
$M_{\Sigma}$ (heavy adjoint scalar multiplet). At issue is how to handle
these thresholds.

In general, thresholds are included in the evolution of the
gauge couplings by a step--function, given logarithmic corrections. For
light thresholds, instead of using specific susy masses one can  used an
effective scale $M_{susy}$ to summarize the effect of the degenerate spectrum
\cite{tsusy}. Corrections due to heavy thresholds in general increase
$M_{susy}$, so that taking a null correction will give a lower
bound on  $M_{susy}$.  Nevertheless, heavy threshold corrections are
constrained
if one takes into account  limits on proton decay via dimension-five operators
(allowed in SGUT's, via the $\Phi$ field) \cite{pd}, and the theoretical
requirement  that Yukawa couplings of heavy scalars do not blow up below the
Planck scale \cite{hisano}.

 In this paper we use a different approach (different from
the step--function approximation)
to include threshold effects in the study of
SGUT's. In a recent paper \cite{mia} we have defined effective charges
\cite{lynn} for the three gauge couplings, calculated with a mass dependent
subtraction procedure \cite{georgi},  which include a $complete$ dependence
on the masses. With this method, one includes, for example, threshold
contributions due to massive gauge bosons  $W^{\pm}$ and $Z^0$, which are
missed in the step--function approximation. In the
step--function approximation each particle with mass $m
\geq m_Z$ contributes to the running coupling \al{i}($\mu$) with a logarithmic
term  $\ln\frac{m^2}{\mu^2}$; on the other hand, the
mass dependent method gives a more precise description of the physics via a
function, for $both$  $m \geq m_Z$ and $m<m_Z$, which can be approximated by
\cite{georgi,mia}:
\begin{equation}
f(\mu,m)=\ln\frac{m_Z^2+c\,m^2}{\mu^2+c\,m^2} \;,
\label{fun}
\end{equation}
where ``$c$" is a constant of order 1--10.
If $m^2\ll m^2_Z$ we
recover the usual term   $\ln\frac{m_Z^2}{\mu^2}$, and when $m^2\gg \mu^2$
there
are no  contributions: the particle  decouples from the theory \cite{appel}.
The constant ``$c$" is chosen so that it ``matches" the logarithm with the
exact
function for intermediate scales, $m_Z< m <\mu$, and its value depends on the
type of particle running inside the loop. When one has two mass
degenerated fermions or two scalars, ``$c$" can be  obtained
from the leading term of a power expansion of the exact threshold function when
$m^2/\mu^2$ goes to infinity. In other cases, the
coefficient given by the power expansion has to be slightly modified to get a
best fitting for intermediate scales.

 To see the basic features of the mass dependent procedure $versus$ the
step--function, we first study the evolution of the coupling constants
from $m_Z$ to $M_X$ without any reference to heavy thresholds. Thus, we
impose the unification condition:
\begin{equation}
\al{1}(M_X)=\al{2}(M_X)=\al{3}(M_X)=\al{G}\;,
\label{unif1}
\end{equation}
with $M_X\geq 10^{16}\,GeV$ to prevent fast proton decay. The
effective charges \al{3}($M_X$), \al{2}($M_X$), \al{1}($M_X$) calculated at
one--loop order are given by:
\begin{equation}
\al{i}(M_X)= \al{i}(m_Z) + \frac{1}{4 \pi} \sum_k b_i^{(k)}
f^{(k)}(M_X,m_k) \;,
\end{equation}
where we sum over all the particles in the SSM, and $f^{(k)}(M_X,m_k)$ is
given by (\ref{fun}). The values of  $m_W$ and $m_Z$ are well known, and we
have
as arbitrary mass parameters: $m_t$ (top quark)\footnote{ The remaining
fermions in the Standard Model are taken as massless.}, and $m_h$ (light Higgs
particle of the SM); $m_{\tilde{w}}$ (winos) and $m_{\tilde{g}}$ (gluinos);
$m_{\tilde{q}}$ (squarks) and $m_{\tilde{l_L}}$, $m_{\tilde{l_R}}$
(sleptons left and right); $m_{\mu}$ (higgsinos); $m_{H}$, $m_+$ and
$m_a$ (scalar masses from the second Higgs doublet needed by susy).
For the susy masses we take a simplified parametrization in terms of
$m_{1/2}$ and $\xi_0=(m_0/m_{1/2})^2$, neglecting the mixing between
winos and higgsinos, and the s--top left and right:
\begin{eqnarray*}
m_{\tilde{w}} &=& m_{1/2}\;\;\;\;,\;\; m_{\tilde{g}}= 3\,m_{1/2}
\\ m_{\tilde{q}} &=&  m_{1/2}\sqrt{7+\xi_0}\;\;\;,\;\;
\mbox{\rm (for all squarks, including s--top)} \\
m_{\tilde{l_L}} &=&  m_{1/2}\sqrt{0.5+\xi_0}\;\;\;, \;\;
m_{\tilde{l_R}}= m_{1/2}\sqrt{0.15+\xi_0}\;\;;
\end{eqnarray*}
we also take $m_{+}\simeq m_a\simeq m_H$. Thus, we are left with
six arbitrary parameters: $m_t$, $m_h$, $m_{1/2}$, $\xi_0$, $m_{\mu}$ and
$m_H$. These parameters have lower bounds \cite{m12} derived from experimental
searches for the top, Higgs and susy particles:
\begin{eqnarray*}
m_t &\geq& 91\; GeV \;\;\;,\;\; m_h \geq 60\; GeV\;\; \cite{mh}\\
m_{1/2} &\geq& 45\; GeV \;.
\end{eqnarray*}
Moreover, perturbative bounds on Yukawa top couplings and quartic Higgs
couplings yield the theoretical upper bound $m_t,\,m_h \leq 200\; GeV$
\cite{cabbibo}; and no extreme fine--tunning on the susy parameters
gives $m_{1/2},\,m_0 \leq 1\; TeV$.

  We also need the initial values \al{i}($m_Z$). The values for
\al{1}($m_Z$) and \al{2}($m_Z$) are derived from the experimental data on
$\sin^2\theta$ and \al{e},
$$
\al{e}(m_Z)=127.9 \pm 0.2\;\;\cite{ale},\;\;
\sin^2\theta(m_Z)= 0.2327\pm 0.0007\;\;\cite{sin},
$$
but for $\alpha_3(m_Z)$ there is no agreement between different
measurements \cite{data3}: The latest LEP data average to
$\alpha_3(m_Z)=0.122$, while data from low energy measurements
average to  $\alpha_3(m_Z)=0.109$. Because of  this discrepancy we will
not take  $\alpha_3(m_Z)$ as the initial data, instead we will
$derive$ it from the
unification condition. In this way susy masses can be bounded by requiring
that $\alpha_3(m_Z)$ be in the range (0.108, 0.125).

In Fig.\ (1) we have plotted $\alpha_3^{-1}(m_Z)$ versus
$\log_{10}(m_{1/2})$, for different values of the remaining free parameters,
and including the experimental and theoretical constraints on
$m_{1/2}$, $\alpha_3^{-1}$ and $M_X$ mentioned above:
\begin{equation}
m_{1/2} > 45\; GeV \;\;,\;\;\; M_X > 10^{16}\;GeV\;\;,\;\;\;
 0.108 \leq \alpha_3(m_Z) \leq 0.125 \;.
\end{equation}
Since $m_{\mu}$, $m_H$ contribute with the same sign, we have simply
taken $m_{\mu}=m_H$ in order to check the basic features of the mass
dependent method. Furthermore, since both  $\alpha_3^{-1}(m_Z)$ and $M_X$
depend very slightly on $\xi_0$, for this plot we have fixed $\xi_0=1$.
Varying the susy masses, we observe a trend similar to what happens when  using
a
step--function: the higher the susy masses, the higher $\alpha_3^{-1}(m_Z)$
and the lower $M_X$.
However, while with the step--function (Fig. 2) one obtains the
limit $\alpha_3(m_Z)< 0.116$ (this procedure does not distinguish
masses lower than $m_Z$),
with the mass dependent procedure  the full range
of $\alpha_3^{-1}(m_Z)$ can be covered for adequate values of the  mass
parameters. Thus the limiting values
$\alpha_3(m_Z)=0.125$ and $M_X=10^{16}$ give us a lower bound on
$m_{\mu}=m_H$ and an upper limit on $m_{1/2}$. These bounds depend on
$m_t,\; m_h$ and $\xi_0$; both of them decrease with $m_t$ and $m_h$,
while the bound on $m_{1/2}$ increases with $\xi_0$ and the bound on
$m_{\mu}$ decreases, remaining practically  unchanged for
$\xi_0 \geq 10^4$. Therefore,
if we allow a maximum difference of two orders of magnitude between $m_0$
and  $m_{1/2}$, we get the following upper bounds
(using the central values of $\alpha_e$
and  $\sin_{\theta}^2$)
\FL
\begin{eqnarray}
m_{1/2}&\leq& 2.5\, TeV \;\;\;\;
(m_h=60\,GeV\,,\;\;m_t=91\,GeV) \;,\nonumber \\
m_{\mu}&\geq& 338\, GeV\;\;\;
(m_h=m_t=200\,GeV\,,\;\;m_{\mu}=m_H) \;.
\end{eqnarray}

 The one--loop calculation already shows the differences between the
step--function  and the mass dependent procedure. The first favors
$\alpha_3(m_Z)$ to be in  the range of the low energy experimental data;
in addition, the lower data $excludes$
 susy masses greater $1\;TeV$. On the other hand, when we include a more
precise
treatment of thresholds, the naturalness bound of $1\; TeV$ $favors$
 $\alpha_3(m_Z)$ to be in the range of the last LEP data.

When we improve the accuracy, calculating at two--loop order, the value of
$\alpha_3(m_Z)$ diminishes by about 10\%. In the case of the step--function,
this effect drives the upper bound obtained at one--loop towards the
experimental upper value, while the naturalness bound on susy masses gives a
lower  bound $\alpha_3(m_Z)\ge 0.118$\footnote{ As was pointed in
Ref.\cite{egm1},  this bound is consistent
with the one obtained including the impact of the evolution of gaugino masses
(EGM) in the step--function approximation \cite{egm2}; the lower bound on
$\alpha_3(m_Z)$ (upper bound on $\alpha_3^{-1}(m_Z)$) comes from taking
$m_{1/2}\approx 45\, GeV$, for which the EGM effect is small. This does not
occur with the ``mass dependent" effect, which gives differences of order 13\%,
no matter the region of masses we take.}.  With the mass dependent procedure, a
10\%  decrease in the one--loop result puts the values of $\alpha_3^{-1}(m_Z)$
obtained with $m_{\mu}=m_H=1\; TeV$ $away$ from the experimental band, being
now
the  lower bound on $m_{\mu}=m_H$ around $10\;TeV$. In
Fig. (3) we have plotted the upper  bound on $m_{1/2}$ and the lower bound
on $m_{\mu}=m_H$ obtained at  two--loop order. The behavior of these
 bounds differs
slightly from  one--loop: the upper bound on $m_{1/2}$ is
a maximum for $\xi_0\simeq 60$  (the value plotted), nearly
independently of $m_t$ and $m_h$, and the lower  bound on $m_\mu$
decreases even for $\xi_0>10^4$. We would need  $\xi_0>10^{10}$ to have
$m_{\mu}$ in the range of $TeV$, and then squark and  slepton masses of
order of $PeV$. Thus, this simple scenario of perturbative  unification at
two--loop order  is $not$  $compatible$ with experimental data on
$\alpha_3(m_Z)$  and the theoretical naturalness bound on the susy
spectrum. We need  at least a heavy higgsino or a  heavy Higgs (or both
of them) beyond this bound.  We also note that the remaining susy masses
(gauginos, squarks and sleptons) are allowed to have values below $1\;TeV$.

Up to now we have not included the effects of heavy thresholds,
but a correct picture of
perturbative unification needs to include them. Thus, the unification
condition (\ref{unif1}) reverts to:
\begin{equation}
4 \pi \alpha_i^{-1}(\mu)=4 \pi \alpha_G^{-1}(\mu)+\lambda_i(\mu,M_j)\;,
\label{unif2}
\end{equation}
where $M_j$ are the heavy masses, and $\mu$ is a mass scale satisfying
$m_i \ll \mu \ll M_j$, $i.e.$, it is far away from both light and heavy
thresholds \cite{das,hall}. The unification condition is obtained taking into
account the decoupling of the heavy degrees of freedom  from the low energy
theory;
integrating out  these fields from the action one gets an effective
field theory in terms  of low energy parameters
($\alpha_i,\;m_i$), which are related to  the high energy parameters
($\alpha_G,\;M_i$) through Eq. (\ref{unif2}).  The functions $\lambda_i$ at
one--loop include logarithmic terms due to heavy degrees of freedom,
and constant terms due to light degrees of freedom, \cite{das}:
\begin{eqnarray}
\lambda_i(\mu,M_j)&=& \lambda_i^{(l)}+\lambda_i^{(H)}(\mu,M_j) \;\;,\\
\lambda_i^{(l)}&=& c_i\left(\frac{10s}{3\sqrt{3}}-\frac{76}{9}\right)+
                   T_h^i \left(\frac{8}{9}\right)+
                   T_f^i\left(\frac{10}{9}\right)\;,\\
\lambda_i^{(H)}(\mu,M_j)&=& 7 \bar{c_i}\ln\frac{M_V}{\mu}-
                   \frac{2}{3} T_H^i \ln\frac{M_H}{\mu}-
                   \frac{4}{3} T_F^i\ln\frac{M_F}{\mu} \;.
\end{eqnarray}
$$(s=2.029884...)$$
Here $h,\,f$ refer to light scalars and fermions respectively, and
$H,\,F$ to the  heavy ones; $c_i=C_2(G_i)$ for $G_i=SU(3),\;SU(2),\;U(1)$,
and  $\bar{c_i}=C_2(G)-C_2(G_i)$; and $T_a^i$ are representation--dependent
coefficients\footnote{ The constant term $-\bar{c_i}/3$ is not present in
$\lambda_i^H$ when one uses the $\overline{DR}$ subtraction procedure
\cite{dr},
as it occurs in supersymmetric theories \cite{kounnas}.}.
 Thus, if we make the reasonable assumption that the members of each heavy
supermultiplet are degenerate we get: \begin{eqnarray}
\lambda_1(\mu,M_j)&=&\frac{66}{5}+\frac{96}{5}\ln\frac{M_V}{\mu}-
                    \frac{4}{5}\ln\frac{M_{\Phi}}{M_V}-
	                   \frac{20}{3}\ln\frac{M_{\Sigma}}{M_V} \;,\\
\lambda_2(\mu,M_j)&=&\frac{20 s}{3\sqrt{3}}-
                    \frac{2}{3}+8\ln\frac{M_V}{\mu}-
	                   8\ln\frac{M_{\Sigma}}{M_V} \;,\\
\lambda_3(\mu,M_j)&=&\frac{10 s}{\sqrt{3}}-10-
                     \ln\frac{M_{\Phi}}{M_V}-
	                    \frac{26}{3}\ln\frac{M_{\Sigma}}{M_V} \;.
\end{eqnarray}

The couplings $\alpha_i^{-1}(\mu)$ in Eq.(\ref{unif2}) are the same as
given by (3), where $M_X$ is now replaced by $\mu\ll M_X$.
Since heavy  masses are typically of order $10^{16}\, GeV$, and light masses
are expected to be less than $1\, TeV$, we  choose $\mu=10^7\,GeV$.
Eliminating  $\alpha_G^{-1}(\mu)$ from (\ref{unif2}), we obtain
$\alpha_3^{-1}(m_Z)$   and $\ln M_V$:
\begin{eqnarray}
\alpha_3^{-1}(m_Z)&=& \frac{1}{2}\left(3\alpha_2^{-1}(m_Z)-
  \alpha_1^{-1}(m_Z)\right)+\frac{1}{8\pi}F_3(m_i,\mu)-
   \frac{3}{5\pi}\ln M_{\Phi}-\frac{3}{5\pi}\;,
\label{alpha3}\\
\ln M_V &=&\frac{3\pi}{8}\left(3\alpha_1^{-1}(m_Z)-
  \alpha_2^{-1}(m_Z)\right)+\frac{3}{32}F_V(m_i,\mu) \nonumber \\
  & & -\frac{3}{40}\ln M_{\Phi}-\frac{1}{8}\ln M_{\Sigma}
      -\frac{13}{10}+\frac{5s}{8\sqrt{3}}\;.
\label{lnv}
\end{eqnarray}
The dependence of $\alpha_3^{-1}(m_Z)$ and $\ln M_V$ with the light masses
(given by the functions $F_3(m_i,\mu)$, $F_V(m_i,\mu)$) is qualitatively the
same as before, $i.e.$, $\alpha_3^{-1}(m_Z)$ increases with the susy
mass parameters while $\ln M_V$ decreases. We now focus our attention on
the heavy mass parameters $M_{\Phi}$ and $M_{\Sigma}$. We see from
Eq.\ (\ref{alpha3}) that
$\alpha_3^{-1}(m_Z)$  depends only on $M_{\Phi}$, so the limits on
$M_{\Phi}$ will put a bound  on $\alpha_3(m_Z)$. The lower bound comes
from the experimental
limits on proton decay via dimension--five operators. Using a chiral
Lagrangian technique, the lifetime obtained for the dominant mode
is \cite{hisano}: \begin{equation}
\tau(p\rightarrow K^+\bar{\nu}_{\mu})=6.9\times10^{31}\left|
 \frac{0.003}{\beta}\frac{\sin2\beta_H}{1+y^{tK}}\frac{M_{\Phi}}{10^{17}}
  \frac{10^{-3}}{f(m_{\tilde{q}}, m_{\tilde{l}},m_{\tilde{w}})}\right|^2
yrs\;,
\end{equation}
where yet three more unknown parameters have popped--in: the hadron matrix
element parameter $\beta$, which ranges from 0.003 to 0.03 $GeV$; the
ratio of vacuum expectation values of two Higgs doublets $\tan\beta_H$;
and the parameter $y^{tK}$, which represents the ratio of the contribution
of the third generation  relative to the second. To allow an $M_{\Phi}$
as low as possible, we take $\beta=0.003,\;\sin2\beta_H=1$, and\footnote{
In this case we take the simplest
choice, because of our experimental ignorance about the value of $y^{tK}$. In
fact, $y^{tK}$ could be negative, giving $|1 + y^{tK}| < 1$.}
$\left|1+y^{tK}\right|=1$. The experimental limit
for this mode is $\tau(p\rightarrow K^+\bar{\nu}_{\mu}) > 1.0\times 10^{32}yrs$
\cite{data}, so we get:
\begin{equation} M_{\Phi} > 1.2\times 10^{20}
f(m_{\tilde{q}}, m_{\tilde{l}},m_{\tilde{w}})= M_{\Phi}^{min}\;.
\label{phimin}
\end{equation}
The function $f(m_{\tilde{q}}, m_{\tilde{l}},m_{\tilde{w}})$, with the
parametrization we have adopted for the susy masses, is given by:
\FL
\begin{equation}
f(m_{\tilde{q}},m_{\tilde{l}},m_{\tilde{w}})=\frac{1}{6.5m_{1/2}}
\left(\frac{\xi_0+13.5}{\xi_0+6}\ln (\xi_0+7)-
      \frac{\xi_0+0.5}{\xi_0-0.5}\ln (\xi_0+0.5)\right)\;,
\end{equation}
and therefore the lower bound only depends on $m_{1/2}$ and $\xi_0$,
decreasing with $m_{1/2}$ and  $m_0$.

On the other hand, both $M_{\Phi}$ and $M_{\Sigma}$ can be bounded from
above by requiring that the Yukawa couplings involving these fields do not
blow up below the Planck scale \cite{hisano}. This leads to
$M_{\Phi}<2M_V$, $M_{\Sigma}<1.8M_V$, and from (\ref{lnv}) we get the
upper bound on $M_{\Phi}$ in terms of the light masses as well as
$M_{\Sigma}$:
\begin{eqnarray}
\ln M_{\Phi}&<&
\frac{15\pi}{37}\left(\alpha_1^{-1}(m_Z)-\alpha_2^{-1}(m_Z)\right)
  +\frac{15}{148}F_V(m_i,\mu) \nonumber \\
  & &-\frac{5}{37}\ln M_{\Sigma}
  -\frac{52}{37}+\frac{200s}{111\sqrt{3}}+\frac{40}{37}\ln 2=\ln M_{\Phi}^{max}
\;.
\label{phimax}
\end{eqnarray}
With $M_{\Phi}^{min}(m_{1/2},\xi_0)$, $M_{\Phi}^{max}(m_i,M_{\Sigma})$ we
get  upper and  lower bounds on $\alpha_3^{-1}(m_Z)$, which have to
be within the range of experimental data. Therefore, we can $play$ with
the expressions (\ref{alpha3}), (\ref{phimin}), (\ref{phimax}), and the
limits on the susy masses trying to check whether or not
the perturbative unification
scenario is compatible with $all$ the constraints.

In order to compare, we first examine the case of the step--function
approximation for $F_3(m_i,\mu)$ and $F_V(m_i,\mu)$. With the constraint
of susy masses below $1\,TeV$ there is no problem in having $\alpha_3$
inside its experimental range. As it can be seen in Fig.(4), the upper
limit on $\xi_0$ gives us the minimum allowed value  for $M_{\Phi}$ and
 the maximum  for $M_{\Sigma}$; furthermore, the lower limit on $M_{\Sigma}$
would give us the lower allowed value for $\xi_0$, which is just reached
for the maximum allowed values of $m_{1/2}$ and $m_0$, and the upper one for
$M_{\Phi}$. In principle, we do not have any constraint on the lower
bound for $M_{\Sigma}$, except the requirement that there is no large
splitting between the heavy masses. As $M_{\Phi}$, $M_V$ will be around
$10^{16}\,GeV$, we use $M_{\Sigma} \geq 10^{13}\,GeV$ to give the results.
Increasing $M_{\Sigma}$  reduces both $M_{\Phi}^{max}$ and
$\alpha_3^{max}$. These results are summarized in Table I, where we have
considered limiting susy masses of $1\,TeV$ and $2\,TeV$ respectively.
For both examples, the allowed range on $\alpha_3(m_Z)$ is inside
the experimental
bounds; raising the limiting mass enlarges it, the same happens with
the allowed range for $M_{\Phi}$ and $m_{1/2}$.
It is seen  that if we require all light masses below $1\,TeV$, and
$M_{\Sigma} \geq 10^{13}\,GeV$, we get $m_{1/2}<m_Z$, near  it lower
experimental bound, and also strong constraints for $m_0$, $M_{\Phi}$,
$M_{\Sigma}$, so that practically,  $m_0\approx 1\,TeV$,
$M_{\Phi}\approx 10^{16.6}\, GeV$, and $M_{\Sigma}\approx 10^{13}\,
GeV$. There are no further constraints on $m_{\mu},\, m_H$ except,
of course, the common upper limit we chose for the susy masses.

As we have printed out  without taking into account heavy thresholds,
the mass dependent procedure gives us values of
$\alpha_3^{-1}(m_Z)$ lower than those obtained with the step--function
approximation.  And now, to satisfy the constraints
on $M_{\Phi}$ and  $\alpha_3^{-1}(m_Z)$ we will need to have
$m_0,\,m_{\mu}$ or $m_H$ beyond $1\, TeV$. The lower $\xi_0$
($m_0$), the higher will be $M_{\Phi}$, and higher values of
$m_{\mu},\, m_H$ will be needed  to raise  $\alpha_3^{-1}(m_Z)$
above its minimum bound
(Fig. 5). The results obtained in this case are given in Table II:
there are no solutions if we maintain all susy masses below $1\, TeV$.

We have seen that the use of a mass dependent procedure and the requirement of
having susy masses not too high, favor a value of $\alpha_3^{-1}(m_Z)$
($\alpha_3(m_Z)$)  near
its lower (upper) experimental bound. This is easily understood, since the mass
dependent procedure has the effect of raising the values of
$\alpha_i^{-1}(\mu)$
at high energies, with respect to those obtained with other methods
($\overline{MS}$, step--function), and their different results tend to
merge as susy masses are increased. When the maximum allowed value for
$\alpha_3^{-1}(m_Z)$ is not enough to unify the couplings we will have to
increase the masses; sometimes above its naturalness bound, as it happens in
the
two--loop calculation without heavy thresholds. Improving our scenario of
coupling constant unification with the effect of heavy thresholds does not
improve the situation about the light masses. In principle, taking $M_{\Phi}$
low enough we will get $\alpha_3(m_Z)\leq 0.125$.  But experimental data on
proton decay do not allow to freely decrease the value of $M_{\Phi}$  without
adequately increasing  the squark and slepton masses. Thus, in the end, we
conclude that it would be necessary to have some of the susy light masses,
(either squarks and sleptons, or higgsinos or heavy Higgs),  heavier than
the naturalness bound of $1\, TeV$ commonly taken, $contrary$ to the
results obtained using the simplest treatment of the light thresholds
given by the step--function approximation.

 We do not want to emphasize the numerical results, which depend on
details of the model (such as susy mass parametrization) as well as on the
experimental data.  We would specially like to advert to their dependence on
the
procedure chosen to  compute the thresholds, which becomes relevant when
running
the couplings over such a huge range of scales:  from $m_Z$ to $M_X$ (all light
thresholds are crossed). It is clear that the  step--function  approximation
gets
worse as the number of crossed thresholds begins to proliferate; this is the
case, for example, in susy theories, where the mass dependent procedure  gives
the exact contribution for each massive degree of freedom, independently of
their
number, and avoids the presence of cumulative errors that afflicts threshold
crossing in a less complete treatment. Therefore, in models with intermediate
mass scales or with many heavy--matter degrees of freedom, the use of mass
dependent procedures is mandatory before reaching conclusions on unification
and
allied phenomena.
\pagebreak
\figure{ Values of $\alpha_3^{-1}(m_Z)$, compatible with the unification
condition (Eq.\ \ref{unif1}), calculated with a mass dependent procedure at
1--loop order, for different values of $m_{\mu}=m_H$: $m_Z$, $1\;TeV$,
$10\,TeV$,  $100\;TeV$. Dotted lines are the experimental limits on
$\alpha_3^{-1}(m_Z)$ and $m_{1/2}$; solid lines are for $m_t=m_h=200\;GeV$, and
dashed lines for $m_t=91\; GeV$ and $m_h=60\; GeV$. The straight lines (solid
for $m_t=m_h=200\;GeV$ and dashed for $m_t=91\; GeV$, $m_h=60\; GeV$)
are the upper limit obtained for
$\alpha_3^{-1}(m_Z)$ when imposing $M_X=10^{16}\;GeV$ ($m_{\mu}=m_H$ increase
along these lines from bottom to top). The allowed region for
$\alpha_3^{-1}(m_Z)$ are to the left of the straight lines, and between the
dotted lines,  $8 \leq \alpha_3^{-1}(m_Z) \leq 9.2$ and $m_{1/2} \geq 45\,
GeV$.
\label{md312}}

\figure{ Same as Fig.\ \ref{md312}, but with $\alpha_3^{-1}(m_Z)$ calculated
with the step--function approximation. \label{sf312}}

\figure{ Lower bound on $m_{\mu}=m_H$ (bottom curves), and upper bound on
$m_{1/2}$ (straight lines), calculated with a mass dependent procedure at
2--loop
order. The lower bound on $m_{\mu}$ is obtained for the lower experimental
limit
on  $\alpha_3^{-1}(m_Z)$ and the upper limit on $\xi_0$ (for the plot we choose
$\xi_0=10^4$, see text). The upper bound on $m_{1/2}$ is obtained for the upper
limit on $M_X$, and $\xi_0\simeq 60$. Solid and dashed lines follow the same
convention as Fig.\ \ref{md312} and \ref{sf312}.  \label{mu12}}

\figure{ $\alpha_3^{-1}(m_Z)$ versus $\log_{10}(M_{\Phi})$ at 1--loop order
with
the step--function approximation, for limiting susy masses of $1\;TeV$,
$m_{1/2}=45\;GeV$ (solid lines) and $m_{1/2}^{max}=53\;GeV$ (dashed lines),
and satisfying different constraints: ($i$) The less slope bottom lines are
obtained with $M_{\Phi}^{min}(m_{1/2},\xi_0)$ (Eq.\ \ref{phimin}), and fixing
$m_{\mu}=m_H=m_Z$; $\xi_0$ increases along these lines, from right to left. We
have marked the points $\xi_0^{max}(m_{1/2}^{min}=45\; GeV)$, and
$\xi_0^{min}(m_{1/2}^{max}=53 \;GeV)$. Dotted lines set $M_{\Phi}^{min}$ and
$M_{\Phi}^{max}$.
($ii$) The most slope lines are obtained with
$M_{\Phi}^{max}(m_{1/2},\xi_0,m_{\mu})$ (Eq.\ \ref{phimax}), with $\xi_0$ fixed
(really depend very slightly on this parameter), $m_{\mu}=m_H$ increasing from
bottom to top, and limiting values for $M_{\Sigma}$,
$M_{\Sigma}^{min}=10^{13}\; GeV$ to the right, and
$M_{\Sigma}^{max}=10^{13.4}\;
GeV$ to the left. \label{sf3phi}}

\figure{ Same as Fig.\ \ref{sf3phi}, now with a mass dependent procedure and
limiting susy masses of $2\;TeV$ (there is no solution for $1\;TeV$);
$m_{1/2}^{min}=45\;GeV$ in solid lines and $m_{1/2}^{max}=86\;GeV$ in dashed
lines. Dotted lines for $M_{\Phi}^{min}$, $M_{\Phi}^{max}$, and lower
experimental value for $\alpha_3^{-1}(m_Z)$ ($\alpha_3^{-1}(m_Z) \geq 8$) .
\label{sm3phi}}
\pagebreak
\begin{table}
  \begin{tabular}{|c|c|c|c|c||c|c|c|c|c|c|}
$m_0^{max}$ & $\xi_0^{max}$&$M_{\Phi}^{min}$&
$M_{\Sigma}^{max}$&$\alpha_3^{min}$
&$m_{1/2}^{max}$&$\xi_0^{min}$&$m_0^{min}$&$M_{\Sigma}^{min}$&
$M_{\Phi}^{max}$&
$\alpha_3^{max}$
 \\
\tableline
1000 & 494 & $10^{16.58}$&$10^{13.4}$&0.112&53&357&936&$10^{13}$&$10^{16.63}$&
0.119
\\
2000 &1975 & $10^{16.06}$&$10^{16.2}$&0.108&221&82&936&$10^{13}$&$10^{16.63}$&
0.122
\\
  \end{tabular}
\caption[]{1--Loop with Step--function$^a$}
\tablenotes{$^{\rm a}$ Mass values in $GeV$}
\end{table}

\begin{table}
  \begin{tabular}{|c|c|c|c|c|c||c|c|c|c|c|}
$m_0^{max}$ & $\xi_0^{max}$&$m_{\mu}^{min}$&$M_{\Phi}^{min}$&
$M_{\Sigma}^{max}$&
$\alpha_3^{min}$&
$m_{1/2}^{max}$&$\xi_0^{min}$&$M_{\Sigma}^{min}$& $M_{\Phi}^{max}$&
$\alpha_3^{max}$
\\
\tableline
1492 & 1100&1492 & $10^{16.29}$ & $10^{16.34}$ & 0.125 & 45 &1100 &$10^{13}$&
$10^{16.29}$ & 0.125
\\
2000 &1975 & 805 & $10^{16.06}$ & $10^{16.37}$ & 0.123 & 86 & 540 &$10^{13}$&
$10^{16.39}$ & 0.125
\\
3000 &4444 & 358 & $10^{15.75}$ & $10^{16.40}$ & 0.120 &333 &  81 &$10^{13}$&
$10^{16.54}$ & 0.125
\\
4000 &7901 & 179 & $10^{15.53}$ & $10^{16.41}$ & 0.118 &4000&   0 &$10^{13}$&
$10^{16.64}$ & 0.125
\\
5000 &12348& 106 & $10^{15.35}$ & $10^{16.42}$ & 0.116 &5000&   0 &$10^{13}$&
$10^{16.72}$ & 0.125
\\
6084 &18278&$m_Z$& $10^{15.20}$ & $10^{16.44}$ & 0.115 &6084&   0 &$10^{13}$&
$10^{16.79}$ & 0.125
\\
  \end{tabular}
\caption[]{1--Loop with MDSP$^a$}
\tablenotes{$^{\rm a}$ Mass values in $GeV$}
\end{table}
\end{document}